\documentclass[aps,prl,twocolumn,showpacs,superscriptaddress]{revtex4}
\usepackage{graphicx}
\begin{document}


\title{Muon spin relaxation studies of magnetic order and 
superfluid density in antiferromagnetic NdOFeAs, BaFe$_{2}$As$_{2}$ and 
superconducting (Ba,K)Fe$_{2}$As$_{2}$}
     \author{A.~A.~Aczel}
     \affiliation{Department of Physics and Astronomy, McMaster University, Hamilton, Ontario L8S 4M1, Canada}
     \author{E.~Baggio-Saitovitch}
     \affiliation{Centro Brasilieiro de Pesquisas Fisicas, Rua Xavier Sigaud 150 Urca, CEP 22290-180
Rio de Janeiro, Brazil}
     \author{S.~L.~Budko}
     \affiliation{Department of Physics and Astronomy and Ames Laboratory, Iowa State University, Ames, Iowa 50011, USA}     
     \author{P.C.~Canfield}
     \affiliation{Department of Physics and Astronomy and Ames Laboratory, Iowa State University, Ames, Iowa 50011, USA} 
     \author{J.~P.~Carlo} 
     \affiliation{Department of Physics, Columbia University, New York, New York 10027, USA}
     \author{G.~F.~Chen}
     \affiliation{Beijing National Laboratory for Condensed Matter Physics, Institute of Physics,
Chinese Academy of Sciences, Beijing 100080, Peoples Republic of China}
     \author{Pengcheng~Dai}
     \affiliation{Department of Physics and Astronomy, University of Tennessee, Knoxville, Tennessee 37996, USA}
     \author{T.~Goko}
     \affiliation{Department of Physics, Columbia University, New York, New York 10027, USA}
     \affiliation{Department of Physics and Astronomy, McMaster University, Hamilton, Ontario L8S 4M1, Canada}
     \affiliation{TRIUMF, 4004 Wesbrook Mall, Vancouver, B.C., V6T 2A3, Canada} 
     \author{W.~Z.~Hu}
     \affiliation{Beijing National Laboratory for Condensed Matter Physics, Institute of Physics,
Chinese Academy of Sciences, Beijing 100080, Peoples Republic of China}
     \author{G.~M.~Luke}
     \affiliation{Department of Physics and Astronomy, McMaster University, Hamilton, Ontario L8S 4M1, Canada}
     \author{J.~L.~Luo}
     \affiliation{Beijing National Laboratory for Condensed Matter Physics, Institute of Physics,
Chinese Academy of Sciences, Beijing 100080, Peoples Republic of China}
     \author{N.~Ni}
     \affiliation{Department of Physics and Astronomy and Ames Laboratory, Iowa State University, Ames, Iowa 50011, USA}  
     \author{D.~R.~Sanchez-Candela}
     \affiliation{Centro Brasilieiro de Pesquisas Fisicas, Rua Xavier Sigaud 150 Urca, CEP 22290-180
Rio de Janeiro, Brazil}
     \author{F.~F.~Tafti}
     \affiliation{Department of Physics, University of Toronto, 60 St. George Street,
Toronto, Ontario, Canada M5S 1A7}
     \author{N.~L.~Wang}
     \affiliation{Beijing National Laboratory for Condensed Matter Physics, Institute of Physics,
Chinese Academy of Sciences, Beijing 100080, Peoples Republic of China}
     \author{T.~J.~Williams}
     \affiliation{Department of Physics and Astronomy, McMaster University, Hamilton, Ontario L8S 4M1, Canada}
     \author{W.~Yu}
     \affiliation{Department of Physics and Astronomy, McMaster University, Hamilton, Ontario L8S 4M1, Canada}
     \author{Y.~J.~Uemura}
     \altaffiliation[author to whom correspondences should be addressed: E-mail
tomo@lorentz.phys.columbia.edu]{}
     \affiliation{Department of Physics, Columbia University, New York, New York 10027, USA}
\date{\today}

\begin{abstract}

Zero-field (ZF) muon spin relaxation ($\mu$SR) measurements   
have revealed static commensurate magnetic order of Fe moments in NdOFeAs below $T_{N} \sim 135$ K,
with the ordered moment size nearly equal to that in LaOFeAs, and confirmed similar 
behavior in BaFe$_{2}$As$_{2}$.
In single crystals of superconducting (Ba$_{0.55}$K$_{0.45}$)Fe$_{2}$As$_{2}$, $\mu$SR spectra
indicate static magnetism with incommensurate or short-ranged spin structure in $\sim$ 70 \%\ of 
volume below $T_{N} \sim$ 80 K, coexisting with remaining volume which exhibits superfluid-response  
consistent with nodeless gap below $T_{c}\sim 30$ K. 

\end{abstract}

\maketitle


The discovery of the iron oxypnictide superconductor 
La(O,F)FeAs ($T_{c} \sim$ 26K) \cite{laofeashosono} has triggered an 
unprecedented burst of quality research. Notable progress includes syntheses of materials with 
higher $T_{c}$'s \cite{xhchen,gfchen,zren} containing different
Rare Earth (RE) elements, hole doping \cite{holedopedhhwen},
reduced oxygen \cite{highpressureoxygen}, and superconducting bi-layer AFe$_{2}$As$_{2}$
(A=Ba,Sr,Ca) systems by (A,K) 
substitution \cite{bakfe2as2,srkfe2as2}
or application of pressure \cite{cafe2as2pres}.     
In characterization, one finds discoveries of commensurate 
antiferromagnetism of parent compounds
LaOFeAs \cite{daineutron} and BaFe$_{2}$As$_{2}$ \cite{bafe2as2neutron} by neutron scattering,
nearly linear scaling of superfluid density and $T_{c}$ by muon spin relaxation ($\mu$SR) 
\cite{luetkenscondmat,drewcondmat,carlocondmat,khasanovcondmat}, and
studies of the electronic phase diagram by neutrons \cite{daiphase} and $\mu$SR \cite{luetkensphase},
all of which exhibit remarkable similarities with the cases of cuprate systems.
Very recent success in the fabrication of superconducting (Ba,K)Fe$_{2}$As$_{2}$ and (Sr,K)Fe$_{2}$As$_{2}$
single crystals \cite{canfieldbak,wangsrk} has enabled detailed studies by ARPES \cite{arpes1,arpes2}, 
and STM \cite{stm}.  Quantum oscillation \cite{quantum} has been observed in
undoped crystals of BaFe$_{2}$As$_{2}$.

Several controversial issues, however, remain open to further studies.
Neutron measurements found antiferromagnetic order of Fe moments below T $\sim$ 135 K in 
LaOFeAs \cite{daineutron}
and CeOFeAs \cite{daiphase} with the ordered Fe moment size of 0.36 and 0.8 Bohr magnetons, respectively.
Bos {\it et al.\/} and Qiu {\it et al.\/} \cite{ndofeasneutron}, however, reported absence 
of corresponding antiferromagnetic order
in NdOFeAs. 
Such a drastic dependence on RE
elements is surprising, and has to be re-examined by other magnetic probes.
Reports on pairing symmetry are divided between those favoring a gap with \cite{stm,gapnodes}
or without \cite{nodelessgap,arpes2} nodes.
Extensive characterization of single crystal specimens is very 
important in the early stage of materials development for production of crystals
with improved qualities in the future.  To address these issues, we have performed $\mu$SR measurements on
ceramic specimens of NdOFeAs and BaFe$_{2}$As$_{2}$, as well as on superconducting 
single crystals of (Ba$_{0.55}$K$_{0.45}$)Fe$_{2}$As$_{2}$.  As reported in 
this letter, our results demonstrate antiferromagnetism of the former two systems nearly identical to 
that of LaOFeAs.  In the superconducting single crystals, 
we found coexisting signals due to incommensurate or short-ranged static magnetism
from a partial volume fraction, and to a superfluid response from the remaining volume.

Ceramic specimes of NdOFeAs and BaFe$_{2}$As$_{2}$, 
with dimensions of 8-10 mm in diameter and 1-2 mm thick, were synthesized 
at IOP in Beijing following the methods published elsewhere \cite{wangmono,srkfe2as2}.
Single crystals of (Ba$_{0.55}$K$_{0.45}$)Fe$_{2}$As$_{2}$ were prepared at Ames Lab., Iowa
following the method in ref. \cite{canfieldbak}.  Small crystallites, with a typical size
of 2 mm $\times$ 2 mm $\times$ 0.1 mm, were aligned with their main surface (ab-plane) 
parallel to each other to form a mosaic specimen of 65 mg in weight covering
the area of $\sim$ 7 mm in diameter.
$\mu$SR measurments were performed at TRIUMF, Vancouver, 
with the beam direction perpenducular to the main surface of each specimen,
following a standard method described in refs. \cite{reviewrmp,reviewscot,saviciprb}.

\begin{figure}[t]
\includegraphics[width=3.2in,angle=0]{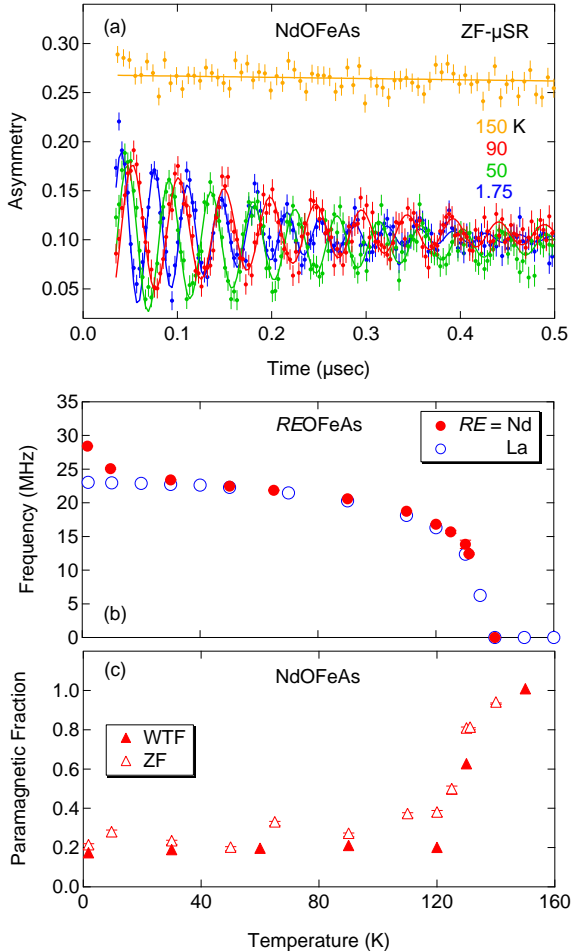}%
\caption{\label{fig1}(color)
(a) Time spectra of Zero-Field (ZF) $\mu$SR in NdOFeAs.  (b) Temperature dependence
of the precession frequency of ZF-$\mu$SR in NdOFeAs (present study) and LaOFeAs \cite{carlocondmat}.
(c) Fraction of muons in a paramagnetic environment, 
estimated from the precession amplitude in weak transverse field (WTF) of 100 G,
as well as from the amplitude of non-oscillating signals in ZF-$\mu$SR spectra.}
\end{figure}

In zero-field (ZF) $\mu$SR measurements of NdOFeAs, long-lived and 
single-component muon spin precession was observed
below $T_{N} \sim$ 135 K, as shown in Fig. 1(a).
The precession amplitude corresponds to the value expected for $\sim$ 70-80 \%\ of muons.
The temperature dependence of the frequency is shown in Fig. 1(b).    
The anomaly of the 
frequency below T = 5 K is presumably due to magnetic ordering of Nd moments,
which was detected by neutron scattering \cite{ndofeasneutron}.
The frequency in NdOFeAs is nearly equal to that 
observed in LaOFeAs \cite{carlocondmat} at $T > 5$ K, indicating that the ordered Fe moment for these two systems 
are  
the same size for temperature regions unaffected by the 
Nd ordering.  Figure 1(c) shows
the fraction of muons in a paramagnetic environment derived from the precessing asymmetry 
observed in
weak transverse field (WTF) of 100 G and also from the ZF-$\mu$SR spectra. 
This figure indicates that a small fraction ($\sim$ 20-30 \%) of muons     
remain in paramagnetic environment even below $T_{N}$, which could either be due to 
a minor paramagnetic volume fraction or to possible cancellation of 
the static local field at corresponding muon sites from symmetry reasons.
Despite this uncertainty,      
the overall results in Fig. 1 clearly demonstrate commensurate antiferromagnetic order
in most of the volume fraction of NdFeAsO below $T_{N}$.  

\begin{figure}[t]
\includegraphics[width=3.2in,angle=0]{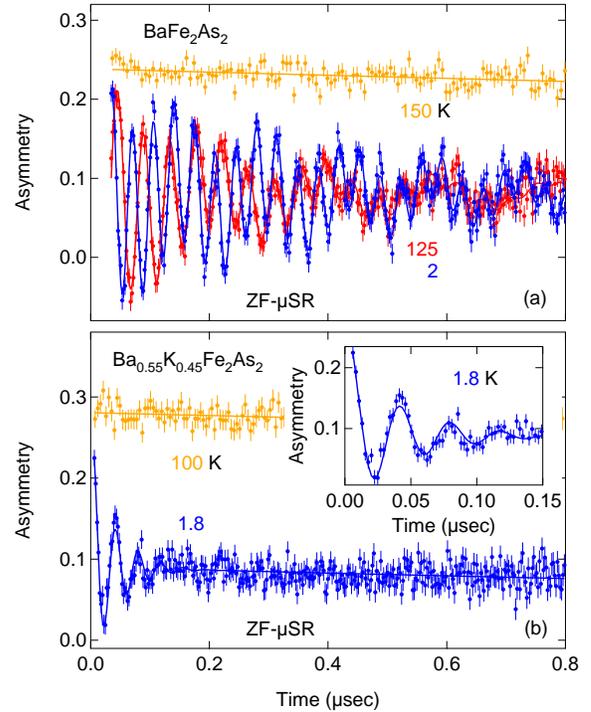}%
\caption{\label{fig2}(color)
Time spectra of ZF-$\mu$SR in (a) ceramic specimen of BaFe$_{2}$As$_{2}$
and in (b) mozaic single crystals of (Ba$_{0.55}$K$_{0.45}$)Fe$_{2}$As$_{2}$.
The inset of (b) compares the observed spectra with Bessel function, 
multiplied by a small exponential damping (solid line).
Long lived oscillation signal in (a) indicates commensurate magnetic order,
while Bessel function shape in (b) suggests possibility of incommensurate
or stripe spin structure.}
\end{figure}

Figure 2(a) shows ZF-$\mu$SR spectra in BaFe$_{2}$As$_{2}$.  Below $T_{N}$ = 140 K,
we find long-lived oscillations with two frequencies: the main frequency at 28.8 MHz
from 80 \% of the muons and the other at 7 MHz from 20 \%, at $T\rightarrow 0$ as shown in 
Fig. 3(a).  The total volume fraction of magnetically ordered region is essentially
100 \%\ below $T_{N}$ (Fig. 3(b)).  Static magnetic order of this system was first 
identified by Rotter et al. \cite{bafe2as2moess} who observed a Moessbauer hyperfine 
field $H_{Moess}$ of 5.47 T at $T \rightarrow 0$
which corresponds to an ordered Fe moment of 0.4 Bohr magnetons. 
In the mono-layer parent system LaOAsFe, the Moessbauer field $H_{Moess}$ 
was 4.86 T \cite{klausscondmat}, while the main frequency in $\mu$SR was 23.0 MHz at $T \rightarrow 0$
\cite{klausscondmat,carlocondmat}.
The ratio of main $\mu$SR frequencies 28.9/23.0 = 1.25 of the bi- and mono-layer systems
is nearly equal to the ratio 5.47/4.86 = 1.13 of $H_{Moess}$, indicating that the effective 
hyperfine coupling constants for muons in these two systems are close to each other within 15 \%.
Combined $\mu$SR, Moessbauer and neutron results demonstrate that antiferromagnetism of 
BaFe$_{2}$As$_{2}$ is nearly identical to that of LaOFeAs and NdOFeAs in terms of $T_{N}$, 
spin structure \cite{daineutron,bafe2as2neutron}, 
ordered moment size, and (nearly full) ordered volume fraction.

\begin{figure}[t]
\includegraphics[width=3.2in,angle=0]{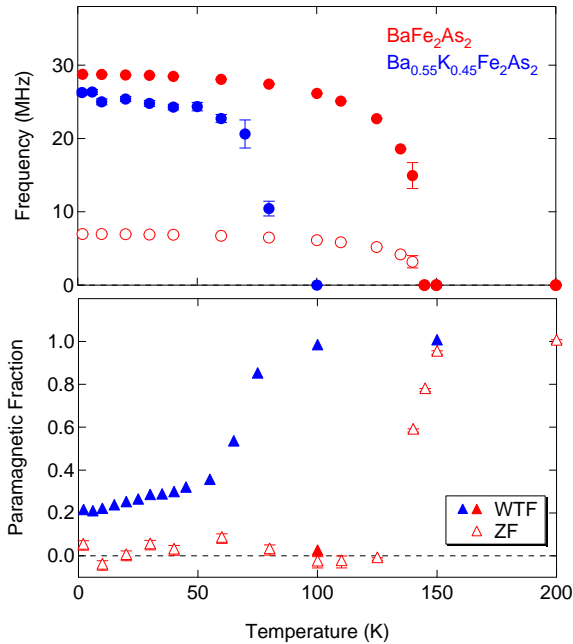}%
\caption{\label{fig3}(color)
(a) Muon spin precession frequency in zero field, and
(b) fraction of signal from muons in paramagnetic / nonmagnetic 
environment, measured in ceramic BaFe$_{2}$As$_{2}$ specimen  
and single crystals of (Ba$_{0.55}$K$_{0.45}$)Fe$_{2}$As$_{2}$.
The paramagnetic fraction was derived as described in the caption
of Fig. 1(c).}    
\end{figure}

In the K-doped single crystals (Ba$_{0.55}$K$_{0.45}$)Fe$_{2}$As$_{2}$, we observed ZF-$\mu$SR 
spectra with a damped precession signal below T = 70 - 80 K,
as shown in Fig. 2(b), only when the initial muon polarization
is perpendicular to the c-axis direction.  This implies an onset of highly inhomogeneous static field, 
parallel to the c-axis, at the muon site(s).  
Although the oscillating spectra can be fit either to a damped cosine
or damped Bessel function, the Bessel-fit (solid line in the inset of Fig. 2(b)) gives the initial phase 
$\phi\sim$ 7 degree at time t = 0 consistent with the experimental condition, unlike the cosine-fit with $\phi\sim$ 
-35 degrees which is significantly off-phase.  This suggests a possibility of incommensurate 
and/or stripe spin structure of the 
ordered region \cite{lpletmtsfprb,lukelbcozf,saviciprb}. Note that a similar damped Bessel signal was
observed also in 3\%\ F doped La(O$_{0.97}$F$_{0.03}$)FeAs \cite{carlocondmat}.   

Figure 3(a) shows the precession frequency in the K doped crystals, which is close to that in 
undoped BaFe$_{2}$As$_{2}$.  This indicates a gradual evolution of spin configuration with K doping.   
The fraction of muons in a paramagnetic environment in Fig. 3(b), derived from the corresponding asymmetry 
of the $\mu$SR spectra in WTF and ZF, demonstrates that the magnetic 
order in (Ba$_{0.55}$K$_{0.45}$)Fe$_{2}$As$_{2}$
is detected by $V_{\mu}$ = 60-80 \%\ of the muons below $T\sim$ 60 K, which implies 
the volume fraction $V_{M} \sim$ 50-70 \%\ of the region with static magnetism \cite{saviciprb}.

\begin{figure}[t]
\includegraphics[width=3.2in,angle=0]{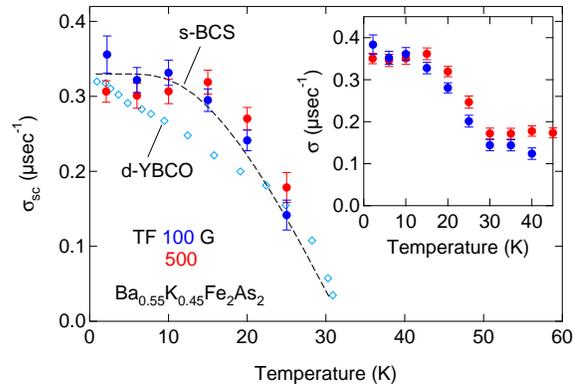}%
\caption{\label{fig4}(color) 
The relaxation rate $\sigma_{sc}$ due to superconductivity in single crystals of 
(Ba$_{0.55}$K$_{0.45}$)Fe$_{2}$As$_{2}$, in transverse field of 100 and 500 G parallel
to the c-axis, obtained after quadratically subtracting the background relaxation 
due to nuclear dipolar fields from the observed Gaussian
relaxation rate $\sigma$ shown in the inset.  The temperature dependence 
is compared to BCS theory for isotropic s-wave pairing (broken line) and 
scaled results from YBCO cuprate superconductor \cite{reviewrmp} (small open
diamond symbols).}
\end{figure}

The remaining paramagnetic signal allowed measurements of the superfluid response below $T_{c}\sim$ 30 K
in (Ba$_{0.55}$K$_{0.45}$)Fe$_{2}$As$_{2}$ single crystals.  The inset of Fig. 4 shows the muon spin relaxation 
rate $\sigma$ observed in transverse external fields of 100 and 500 G parallel to the 
c-axis, obtained in a fit to a Gaussian damping
$\exp(-\sigma^{2}t^{2}/2)$.  This relaxation is related to the magnetic field penetration depth $\lambda$
and $n_{s}/m^{*}$ (superconducting carrier density / effective mass) as $\sigma \propto \lambda^{-2}
\propto n_{s}/m^{*}$ \cite{uemuraprl89,uemuraprl91,reviewrmp,reviewscot}.  
Limited counting statistics of data due to small amount of specimens, small paramagnetic
fraction, and slow relaxation prevented us from fitting the damping to a more complicated line shapes calculated for an 
Abrikosov vortex lattice.  The relaxation rate $\sigma_{sc}$ related to superconductivity has
been obtained by quantratically subtracting the background nuclear dipolar relaxation, 
determined at T = 30 - 40 K, from the observed data.  The temperature dependence of $\sigma_{sc}$ in 
Fig. 4 is compared to the nodeless weak-coupling s-wave BCS curve (broken line) as well as 
to scaled results from
YBCO cuprates \cite{reviewrmp} (open diamonds) representing d-wave gap with line nodes.  The present results
clearly exhibit better agreement with the case of a nodeless isotropic gap.

The observed $\sigma_{sc}(T\rightarrow 0) \sim$ 0.33 $\mu$s$^{-1}$ for crystal specimens 
imply that the superfluid density in the present (Ba$_{0.55}$K$_{0.45}$)Fe$_{2}$As$_{2}$
specimens is about 30 \%\ of that in La(O,F)FeAs with comparable $T_{c}$.  
The small superfluid density and a large volume with static magnetism suggest that
the system is a mixture of regions with static magnetic order and remaining regions with superconductivity.  
This picture is consistent with the results of the 
specific-heat jump reported in ref. \cite{canfieldbak}, as well as STM
studies which found two distinct responses \cite{stm}.

Structural and magnetic phase transition at T = 85 K was found  
in single crystals of undoped BaFe$_{2}$As$_{2}$ prepared with the Sn-flux method \cite{canfieldbak}.
The magnetic order below $T \sim$ 80 K in the present K-doped crystals, prepared
with the same method, suggests a possibility
of a wide spatial spread of K concentrations,
even beyond $\pm$ 7\%\ layer-by-layer spread around 45 \%\ nominal K concentration
found in ref. \cite{canfieldbak}. 
At this moment, however, it is not clear whether the superconducting and magnetically ordered
regions are patterned as a 
microscopic phase separation, similarly to oxygen-overdoped LSCO cuprates
\cite{saviciprb}, or as a segregation of a more macroscopic length scale.

The present studies are made on the first set of superconducting single crystals ever produced
among the bi-layer iron-based systems.  Therefore, the results should be received
with a caution that improved quality / homogeneity / size of the specimen and 
improved statistics of data could possibly alter
the essential message, as has been experienced for the cases of cuprate systems.
Within such limitations, however, the present results indicate: (a) decent superfluid
response of the crystals, (b) possibility of intrinsic microscopic coexistence of 
regions with and without static magnetic order, (c) nodeless energy gap,
which might be intrinsic or be related to scattering of carriers due to 
magnetic volumes, and (d) possibility of involvement of 
incommensurate or stripe spin correlations near the border of antiferromagnetic 
and superconducting states.          

Our $\mu$SR results on NdOFeAs are clearly inconsistent with  
the neutron studies of NdOFeAs \cite{ndofeasneutron} which found an absence of 
corresponding Fe ordering.  Neutron measurments in BaFe$_{2}$As$_{2}$ \cite{bafe2as2neutron} 
and LaOFeAs \cite{daineutron} estimate the size of the ordered Fe moment 
to be 0.87 and 0.36 Bohr mangetons, respectively. 
In contrast, $\mu$SR and Moessbauer studies found a comparable 
moment size in these two systems.  The origin of these disagreements is unclear at this moment.
For the estimate of ordered moment size, however, local probes like $\mu$SR and Moessbauer
generally provide more accurate information than volume-integrated
Bragg-peak intensity of neutron scattering.

In summary, our $\mu$SR measurements of 
Fe-based high-$T_{c}$ systems have revealed 
static magnetic order of NdOFeAs, demonstrated similar magnetic behavior in 
parent compounds of mono- and bi-layer systems 
NdOFeAs, LaOFeAs, and BaFe$_{2}$As$_{2}$, and elucidated coexisting static magnetism
and superconducting responses in single crystals of (Ba$_{0.55}$K$_{0.45}$)Fe$_{2}$As$_{2}$.  

We acknowledge financial support from
US NSF DMR-05-02706 (Material World Network) at Columbia,
US NSF DMR-07-56568 at U. Tennessee Knoxville,
US DOE under contract No. DE-AC02-07CH11358 at Ames,
NSERC and CIFAR (Canada) at McMaster, 
CNPq on MWN-CIAM program at CBPF (Brazil),
and NSFC, CAS, and 973 project of MOST
(China) at IOP, Beijing.
\\

\vfill \eject
\end{document}